\documentclass[aps,prd,twocolumn,a4paper,groupedaddress,showpacs,floatfix]{revtex4-1}
\pdfoutput=1
\usepackage{graphicx}
\usepackage{atlasphysics}
\usepackage{amsmath}

\usepackage[hyperindex,breaklinks,colorlinks]{hyperref}
\hypersetup{linkcolor=blue,citecolor=blue,filecolor=black,urlcolor=blue}

\renewcommand{\pt}{$p_{\rm T}$}

\newcommand{\xF}{$x_{\rm F}$}

\newcommand{\LamDecay}{$\Lambda \to p \pi^-$}
\newcommand{\LamBarDecay}{$\bar\Lambda \to \bar p \pi^+$}

\newcommand{\KsDecay}{$K_S^0 \to \pi^+ \pi^-$}

\newcommand{\Ks}{$K_S^0$}

\renewcommand{\Jpsi}{$J/\psi$}

\newcommand{\Lam}{$\Lambda$}
\newcommand{\LamBar}{$\bar\Lambda$}

\newcommand{\Vz}{$V^0$}

\renewcommand{\sp}{\ \ \,}

\newcommand{\paperTitle}{Measurement of the transverse polarization of $\Lambda$ and $\bar{\Lambda}$ hyperons produced in proton--proton collisions at $\sqrt{s}=7$ TeV using the ATLAS detector}

\newcommand{\paperAbstract}{
The transverse polarization of $\Lambda$ and $\bar\Lambda$ hyperons produced in proton--proton collisions at a center-of-mass energy of 7~TeV is measured. The analysis uses 760~$\mu$b$^{-1}$ of minimum bias data collected by the ATLAS detector at the LHC in the year 2010. The measured transverse polarization averaged over Feynman $x_{\rm F}$ from $5\times 10^{-5}$ to 0.01 and transverse momentum $p_{\rm T}$ from 0.8 to 15~GeV is $-0.010 \pm 0.005({\rm stat}) \pm 0.004({\rm syst})$ for $\Lambda$ and $0.002 \pm 0.006({\rm stat}) \pm 0.004({\rm syst})$ for $\bar\Lambda$. It is also measured as a function of $x_{\rm F}$ and $p_{\rm T}$, but no significant dependence on these variables is observed. Prior to this measurement, the polarization was measured at fixed-target experiments with center-of-mass energies up to about 40~GeV. The ATLAS results are compatible with the extrapolation of a fit from previous measurements to the $x_{\rm F}$ range covered by this measurement.
}

\usepackage{preprintcover}  
\PreprintCoverPaperTitle{\paperTitle}  
\PreprintIdNumber{CERN-PH-EP-2014-258}  
\PreprintCoverAbstract{
  \paperAbstract
}
\PreprintJournalName{Phys. Rev. D}  

\begin{document}


\title{\paperTitle}

\author{The ATLAS Collaboration}
\date{\today}


\begin{abstract}
  \paperAbstract
\end{abstract}


\maketitle

\section{Introduction}
\label{sec:intro}

The transverse polarization of \Lam\ and \LamBar\ hyperons is measured using proton--proton collisions at a center-of-mass energy of 7~TeV, using the data collected by the ATLAS experiment at the Large Hadron Collider (LHC). The \Lam\ hyperons are spin-$1/2$ particles and their polarization is characterized by a polarization vector $\vec P$. Its component, $P$, transverse to the \Lam\ momentum is of interest since for hyperons produced via the strong interaction parity conservation requires that the parallel component is zero. Following the definition used by previous proton--proton and proton--nucleon fixed-target experiments \cite{Bun76, *Hel78, *Sch78, *Hel83, *Yul91, Lun89, Ram94, Fan99, Hera-B} and an experiment at the CERN Intersecting
Storage Rings (ISR) \cite{Erhan1979301}, the polarization is measured in the direction normal to the production plane of the \Lam\ hyperon:
\begin{equation}
  \vec n = \hat p_{\rm beam} \times \vec p, \notag
  \label{eqn:norm}
\end{equation}
where $\hat p_{\rm beam}$ is aligned with the direction of the proton beam and $\vec p$ is the \Lam\ momentum. 

The polarization is measured as a function of the \Lam\ transverse momentum with respect to the beam axis, \pt, and the Feynman-$x$ variable \xF\ defined as $x_{\rm F} = p_{z}/p_{\rm beam}$, where $p_{z}$ is the $z$-component of the \Lam\ momentum and $p_{\rm beam} = 3.5$~TeV is the proton beam momentum. ATLAS uses a right-handed coordinate system with its origin at the nominal interaction point (IP) in the center of the detector and the $z$-axis along the beam pipe. The $x$-axis points from the IP to the center of the LHC ring, and the $y$-axis points upward. Cylindrical coordinates $(r,\phi)$ are used in the transverse plane, $\phi$ being the azimuthal angle around the beam pipe. The pseudorapidity is defined in terms of the polar angle $\theta$ as $\eta=-\ln\tan(\theta/2)$. Since the LHC collides proton--proton beams, the orientation of the $\hat p_{\rm beam}$ direction is arbitrary, which implies the following symmetry of the transverse polarization:
\begin{equation}
  P(-x_{\rm F}) = -P(x_{\rm F}).
  \label{eqn:pol_odd}
\end{equation}
In this analysis, the unit vector $\hat p_{\rm beam}$ is chosen to point in the direction of the $z$-axis of the ATLAS coordinate system for the \Lam\ and \LamBar\ hyperons with positive rapidity and in the opposite direction otherwise. For consistency, \xF\ is treated as positive in both hemispheres. Equation~(\ref{eqn:pol_odd}) implies that the polarization for $x_{\rm F}=0$ is zero. 

The decays \LamDecay\ and \LamBarDecay\ are used to measure the polarization of \Lam\ and \LamBar. In the rest frame of the mother particle, the angle $\theta^*$ between the decay proton (anti-proton) and the direction $\vec n$ follows the probability distribution:
\begin{equation}
  g(t; P) = \frac{1}{2}\left(1+\alpha P t\right),
  \label{eqn:cosThetaTrue}
\end{equation}
where $t \equiv \cos\theta^*$, $\alpha = 0.642 \pm 0.013$ is the world average value of the parity-violating decay asymmetry for the \Lam\ \cite{PDG, Astbury:1975hn, *Cleland:1972fa, *Dauber:1969hg, *Overseth:1967zz, *Cronin:1963zb}. Assuming $CP$ conservation, the value of $\alpha$ for the \LamBar\ decay is of the same magnitude as for the \Lam\ with an opposite sign.

Large polarization of \Lam\ hyperons (up to 30\%) was observed in inclusive proton--proton and proton--nucleon collisions by previous experiments \cite{Bun76, *Hel78, *Sch78, *Hel83, *Yul91, Lun89, Ram94, Fan99, Hera-B, Erhan1979301}. These results are inconsistent with perturbative QCD (pQCD) calculations \cite{Dharmaratna:1996xd} that predict much smaller polarization values. On the other hand, the \LamBar\ polarization was measured to be consistent with zero by all previous experiments \cite{Bun76, *Hel78, *Sch78, *Hel83, *Yul91, Lun89, Ram94, Fan99, Hera-B, Erhan1979301}. Many models \cite{Andersson:1979wj, *Deg81a, *Deg81b, *Szw81, *Fel99, *Boe07, *Siv10, *Anselmino11} have been proposed to explain the origin of hyperon polarization; however, while some models have been successful in explaining the behavior of one member of the hyperon family, no model has been successful in explaining the behavior of all. 

The ATLAS measurement covers a different range in \xF\ and is for a different center-of-mass energy than the previous results. The center-of-mass energy for this measurement is 7~TeV compared to about $ 40$~GeV of the fixed-target experiments \cite{Bun76, *Hel78, *Sch78, *Hel83, *Yul91, Lun89, Ram94, Fan99, Hera-B} and 62~GeV at the ISR collider \cite{Erhan1979301}. The \xF\ coverage of the experiments is different due to the different geometrical acceptances: while the fixed-target experiments covered an $|x_{\rm F}|$ region of roughly 0.01--0.6, in this analysis \Lam\ hyperons are reconstructed with $x_{\rm F}$ ranging from zero to about $0.01$. Furthermore, since most of hyperons are produced at low $|x_{\rm F}|$, in the current data sample enough events are collected only up to $|x_{\rm F}| \approx 0.002$. 

The previous experiments have observed some common features of the \Lam\ polarization:
\begin{itemize}
  \item the magnitude of the \Lam\ polarization increases with \pt\ until it saturates at about 1~GeV;
  \item the magnitude of the \Lam\ polarization decreases with decreasing $|x_{\rm F}|$;
  \item the \Lam\ polarization does not depend strongly on the center-of-mass energy, tested up to $\sqrt{s} \approx 40$~GeV.
\end{itemize}
Although no successful model of \Lam\ polarization exists to date, one can extrapolate the results of the previous experiments into the \xF\ range covered by this analysis assuming no dependence on collision energy (using e.g. the fit to data in Ref.~\cite{Lun89}). Such an extrapolation suggests that a vanishingly small polarization should be seen for small \xF\ at the LHC.

\section{The ATLAS experiment}
\label{sec:atlas}

ATLAS \cite{ATLAS} is a general-purpose detector that covers a large fraction of the solid angle around the IP with layers of tracking detectors, calorimeters, and muon chambers. This measurement makes use of the innermost subsystem called the inner detector (ID). The ID, which serves as a tracking detector, is situated in a magnetic field of about $2$~T generated by a superconducting solenoid. Charged-particle tracks are reconstructed with $p_{\rm T} > 50$~MeV and in the pseudorapidity range of $|\eta| < 2.5$. The ID consists of two types of silicon detectors (pixel and microstrip) and a gas-filled detector with a transition radiation detection capability, called the transition radiation tracker (TRT). The ID forms a cylinder 7~m long with a diameter of 2.3~m. Pixel detectors form three cylindrical layers surrounding the IP and three disks on each side of the detector. Microstrip detectors are arranged in four cylindrical layers (with each module consisting of two silicon wafers) and nine disks in each of the forward regions. The silicon detectors extend to a distance of 50~cm from the beam axis. The TRT consists of multiple layers of gas-filled straw tubes oriented parallel to the beam in the central region and perpendicularly in the forward regions. Typically, three pixel layers and eight microstrip layers (providing four space points) are crossed by each track originating at the collision point. A large number of tracking points (typically 36 per track) are provided by the TRT.  

Events are selected for collection using the minimum bias trigger scintillators (MBTS) \cite{trigger} located between the ID and the calorimeters on both sides of the detector. Each of the scintillators is divided into two rings in pseudorapidity ($2.09 < |\eta| < 2.82$  and $2.82 < |\eta| < 3.84$) and eight sectors in $\phi$. The MBTS trigger selects non-diffractive and diffractive inelastic collision events by detecting forward particle activity in the event. 

\section{Data samples and event selection}
\label{sec:data}

\subsection{Experimental data}

This study uses data collected at the beginning of the year 2010 since they have on average only about 1.07 inelastic proton--proton collisions per bunch-crossing and tracks are reconstructed with a lower \pT\ threshold than used in later runs. The integrated luminosity of the dataset is 760~$\mu$b$^{-1}$. Events must pass the trigger selection requiring at least one hit in either of the two MBTS sides. This trigger has nearly 100\% efficiency for events with more than three tracks \cite{trigger}. Events containing at least one reconstructed collision vertex built from at least three tracks are further analyzed. If there is more than one collision vertex in the event, the one with the largest sum of track $p_{\rm T}^2$ is used as the primary vertex (PV). 

Long-lived two-prong decay candidates (\Vz) are reconstructed by refitting pairs of oppositely charged tracks with a common secondary vertex constraint. The invariant mass of the \Vz\ candidate is calculated using the hypotheses \LamDecay, \LamBarDecay, \KsDecay, and $\gamma\to e^+e^-$ by assigning the mass of the proton, pion, or electron to tracks. Since ATLAS has limited capability to identify protons and pions in the \pT\ range relevant to this analysis, the invariant mass is the main criterion to distinguish between different \Vz\ decays. 

The \Lam\ decay vertex is required to lie within the volume enclosed by the last layer of the silicon tracking detectors, which restricts the transverse decay distance of \Lam\ to about 45~cm. Hyperons that decay outside of this volume (about 15\%) are not reconstructed.

\subsection{Monte Carlo simulation}

Monte Carlo (MC) simulation is used to estimate effects of the detector efficiency and resolution. A sample of 20\,million minimum bias events was generated with {\sc Pythia~6.421} \cite{Pythia} using the ATLAS minimum bias tune (AMBT1) \cite{ATL-PHYS-PUB-2011-008} and MRST2007LO$^{*}$ parton distribution functions \cite{MRST2007LOmod}. In addition, 70~million simulated single-\Lam\ events are combined with the minimum bias sample. The \textsc{Geant}4 package \cite{geant} is used to simulate the propagation of generated particles through the detector \cite{Aad:2010ah}. \textsc{Geant}4 also handles decays of long-lived unstable particles, such as pions, kaons, and \Lam\ hyperons. It decays \Lam\ and \LamBar\ hyperons with a uniform angular distribution, which corresponds to zero polarization. The samples are reconstructed using software with a configuration consistent with the one used for the data. Differences between the reconstruction performance for \Lam\ and \LamBar\ in the minimum bias and single-\Lam\ MC samples are found to be negligible. 

The main differences between \Lam\ and \LamBar\ hyperons are in their production cross sections and in the interaction of the decay proton and anti-proton with the detector material. However, the predicted distributions of the decay angle are consistent between these samples. Therefore, the simulated samples of \Lam\ and \LamBar\ are combined. 

\subsection{Signal selection}

The simulated sample is used to optimize the selection requirements employed to separate samples of \Lam\ and \LamBar\ decays from background. Two types of background are considered. The combinatorial background consists of random combinations of oppositely charged tracks, usually originating at the PV, that mimic a \Lam\ decay. The physics background consists of \Ks\ decays and $\gamma \to e^+e^-$ conversions that are misidentified as \Lam. 

Three main sets of selection criteria are used to address these backgrounds. The first type is aimed at selecting candidates that are reconstructed with good quality by requiring a vertex fit probability greater than $0.05$. The second category includes requirements on the \Lam\ transverse decay distance significance ($L_{xy}/\sigma_{L_{xy}} > 15$) and \Lam\ impact parameter significance ($a_0/\sigma_{a_0} < 3$). The transverse decay distance $L_{xy}$ is defined as a projection of the vector connecting the primary and secondary vertices onto the \Lam\ direction, where only $xy$ vector components are taken into account. The \Lam\ impact parameter $a_0$ is defined as a shortest distance (in three dimensions) between the PV and the line aligned with the \Lam\ momentum. Uncertainties of $L_{xy}$ and $a_0$ are denoted $\sigma_{L_{xy}}$ and $\sigma_{a_0}$, respectively. These two requirements suppress combinatorial backgrounds due to particles originating at the PV and secondary interactions with the detector material. The third type of selection criteria aims to reduce physics background by removing candidate vertices in the \Ks\ invariant mass window ($480 < m_{\pi\pi} < 515$~MeV) and photon conversion candidates ($m_{ee} < 75$~MeV). 

The fiducial phase space for the \Lam\ (\LamBar) candidates is defined by the following requirements: $0.8 < p_{\rm T} < 15$~GeV, $5\times 10^{-5} < x_{\rm F} < 0.01$, and $|\eta| < 2.5$. The majority of reconstructed candidates (90\%) falls into these intervals. The lower bound of the \xF\ range is chosen to remove candidates whose rapidity sign can be mis-measured due to detector resolution. 

Furthermore, only candidates with an invariant mass in the range $1100 < m_{p\pi} < 1135~{\rm MeV}$ are analyzed. The signal region is defined as $1105 < m_{p\pi} < 1127$~MeV and the ranges $1100 < m_{p\pi} < 1105~{\rm MeV}$ and $1127 < m_{p\pi} < 1135~{\rm MeV}$ are referred to as sidebands. 

With these selection criteria, 423\,498 \Lam\ and 378\,237 \LamBar\ candidates are selected in the full mass range in data. The difference between the two numbers is caused by different production cross sections for \Lam\ and \LamBar\ (which is less than 5\% \cite{PhysRevD.85.012001}), different absorption cross sections with the detector material, and differences in the reconstruction efficiencies at low \pt. The average \pT\ of the selected candidates is 1.91~GeV and the average \xF\ is 0.001. No candidate is successfully selected under both the \Lam\ and \LamBar\ hypotheses.

\subsection{Weighting of simulated samples}

The simulated samples are adjusted by event weighting to improve agreement with data (kinematic weight) and to allow assignment of different polarizations to the \Lam\ hyperons (polarization weight). 

The kinematic weight is expressed as a function of \pt\ and $\eta$ of the \Lam\ hyperons and the longitudinal position of the PV, $z_{\rm PV}$. It is calculated as a ratio of data to MC distributions. Since \pt\ and \eta\ are correlated, a two-dimensional weight histogram $w(p_{\rm T}, \eta)$ is used. A single-variable weight histogram $w(z_{\rm PV})$ is used to correct the PV position. The final kinematic weight is expressed as a product, $w(p_{\rm T}, \eta, z_{\rm PV}) = w(p_{\rm T}, \eta)\,w(z_{\rm PV})$. Both data and MC histograms are constructed using events in the signal region. For the data, distributions of events in the sidebands are subtracted from the signal region to approximate signal-only distributions. The signal fraction used for the background subtraction is determined in the mass fit described in Sec.~\ref{sec:mass_fit}. To regularize the weight function, values are linearly interpolated between the bin centers of the weight histograms. In data, the \Lam\ and \LamBar\ samples are kept separate. Therefore, the kinematic weights are computed separately for the observed \Lam\ and \LamBar\ distributions. Data and weighted MC-event distributions of \pT, \eta, rapidity, and transverse decay distance of the hyperons were compared to ensure that the corrected samples describe the data well. In addition, distributions of the number of hits in the silicon detectors and the transverse momenta of the vertex-refitted final-state tracks were also compared between data and simulation and found to be in agreement. 

Since the original MC sample is unpolarized, an event weight of
\begin{equation}
  w_P (t) = 1 + \alpha P t, 
  \label{eqn:wPol}
\end{equation}
where $t=\cos\theta^*$, is applied together with the kinematic weight to include the polarization in the MC sample. The final event weight is expressed as a product of the kinematic and polarization weight. The weight function can be factorized in this way since distributions of \pT, \xF, and the transverse momenta of individual tracks are found to be unaffected by the decay angle reweighting.

\section{Measurement strategy}
\label{sec:strategy}

The polarization is measured by analyzing the angular distribution of the \Lam\ and \LamBar\ decay products. For reconstructed decays, Eq.~(\ref{eqn:cosThetaTrue}) is modified by detector efficiency and resolution effects:
\begin{equation}
  g_{\rm det}(t'; P) \propto \frac{1}{2} \int\limits_{-1}^1 dt\, \left[ \left( 1 + \alpha P t\right) \, \varepsilon(t) \right]\,  R( t', t),
  \label{eqn:cosTheta}
\end{equation}
where $t' \equiv \cos\theta^*_{\rm det}$ is the cosine of the measured decay angle, $\varepsilon(t)$ is the reconstruction efficiency, and $R( t', t)$ is the resolution function, which is convolved with the decay angle distribution. Both the efficiency and the resolution functions are obtained from the MC simulation, as explained later. The typical resolution of the cosine of the \Lam\ decay angle is 0.03.

The method of moments is used to extract the value of the polarization $P$. It exploits the fact that, for any value of $P$, the first moment of Eq.~(\ref{eqn:cosTheta}) can be expressed as a linear combination of the first moments of distributions with polarization $P=0$ and $P=1$: 
\begin{eqnarray}
  E(P) &=& \int\limits_{-1}^{1} dt'\ t'\ g_{\rm det}(t'; P) = \notag\\
       &=& E(0) + \left[ E(1) - E(0)\right]\,P. \notag
  \label{eqn:E}
\end{eqnarray}
The moments $E(0)$ and $E(1)$ can be estimated using MC simulation as averages of the reconstructed decay angle values for samples with polarization 0 and 1. 

To correct for the background contribution, the first moments are calculated separately in the signal and sideband regions. It is assumed, and verified with the MC simulation, that the the first moment of the angular distribution of the background, $E_{\rm bkg}$, is consistent with being independent of $m_{p\pi}$. The value measured in the sidebands is therefore used for the signal region. Given the values of $P$ and $E_{\rm bkg}$, the expected first moment in each of the three mass regions can therefore be expressed as
\begin{eqnarray}
  E^{\rm exp}_i(P, E_{\rm bkg}) &=& f^{\rm sig}_i \big\{ E^{\rm MC}_i(0)  + \notag\\
          &+&  \big[E^{\rm MC}_i(1) - \ E^{\rm MC}_i(0)\big]P \big\} + \notag\\
          &+&  (1-f^{\rm sig}_i) E_{\rm bkg},
  \label{eqn:E_final}
\end{eqnarray}
where $E^{\rm MC}_{i}(0)$ and $E^{\rm MC}_{i}(1)$ are the first moments of the angular distributions for $P=0$ and $P=1$, respectively, estimated using the MC simulation, and $f^{\rm sig}_i$ are the corresponding signal fractions in each of the three mass regions, denoted by index $i = 1, 2, 3$. 

The values of $P$ and $E_{\rm bkg}$ are extracted using a least-squares fit, i.e. by minimizing
\begin{equation}
  \chi^2(P, E_{\rm bkg}) = \sum\limits_{i=1}^{3} \frac{[E_i - E^{\rm exp}_i(P, E_{\rm bkg})]^2}{\sigma_{E_i}^2},
  \label{eqn:chi2}
\end{equation}
where $E_i$ are the first moments of the angular distributions measured in data and $\sigma_{E_i}$ are the corresponding statistical uncertainties. Only $P$ and $E_{\rm bkg}$ are treated as free parameters in the fit. The signal fractions $f^{\rm sig}_i$ and the first moments of the simulated angular distributions $E^{\rm MC}_{i}(0)$ and $E^{\rm MC}_{i}(1)$ are fixed.

\section{Signal fraction extraction}
\label{sec:mass_fit}

Fits to the \Lam\ and \LamBar\ invariant mass distributions are used to extract the signal fraction in the signal region and the two sidebands. A two-component mass probability density function  is used:
\begin{equation}
  \mathcal{M}(m_{p\pi}) = f_{\rm sig} \mathcal{M}_{\rm sig}(m_{p\pi}) + (1-f_{\rm sig}) \mathcal{M}_{\rm bkg}(m_{p\pi}), \notag
  \label{eqn:fit}
\end{equation}
where $\mathcal{M}_{\rm sig}(m_{p\pi})$ and $\mathcal{M}_{\rm bkg}(m_{p\pi})$ denote the signal and background components, respectively. The signal component is defined as a triple asymmetric Gaussian distribution:
\begin{eqnarray}
  \mathcal{M}_{\rm sig}(m_{p\pi}) &=& f_1 \mathcal{G}(m_{p\pi}; m_\Lambda, \sigma_1^{\rm L}, \sigma_1^{\rm R}) + \notag\\
                                  &+& (1-f_1) \Big[ f_2 \mathcal{G}(m_{p\pi}; m_\Lambda, \sigma_2^{\rm L}, \sigma_2^{\rm R}) + \notag\\
                                  &+& (1-f_2) \mathcal{G}(m_{p\pi}; m_\Lambda, \sigma_3^{\rm L}, \sigma_3^{\rm R}) \Big], \notag
  \label{eqn:fit_sig}
\end{eqnarray}
where $\mathcal{G}(m_{p\pi}; m_\Lambda, \sigma^{\rm L}, \sigma^{\rm R})$ is an asymmetric Gaussian function with most probable value $m_\Lambda$, and left and right widths $\sigma^{\rm L}$ and $\sigma^{\rm R}$. The relative contributions of the first and second Gaussian functions are denoted $f_1$ and $f_2$, respectively. The parameters $m_\Lambda$, $\sigma_{1}^{\rm L}$, $\sigma_{2}^{\rm L}$, $\sigma_{3}^{\rm L}$, $\sigma_{1}^{\rm R}$, $\sigma_{2}^{\rm R}$, $\sigma_{3}^{\rm R}$, $f_1$, and $f_2$ are treated as free parameters in the fit. 

\begin{figure}[tb] 
  \centering
  \includegraphics[width=\linewidth]{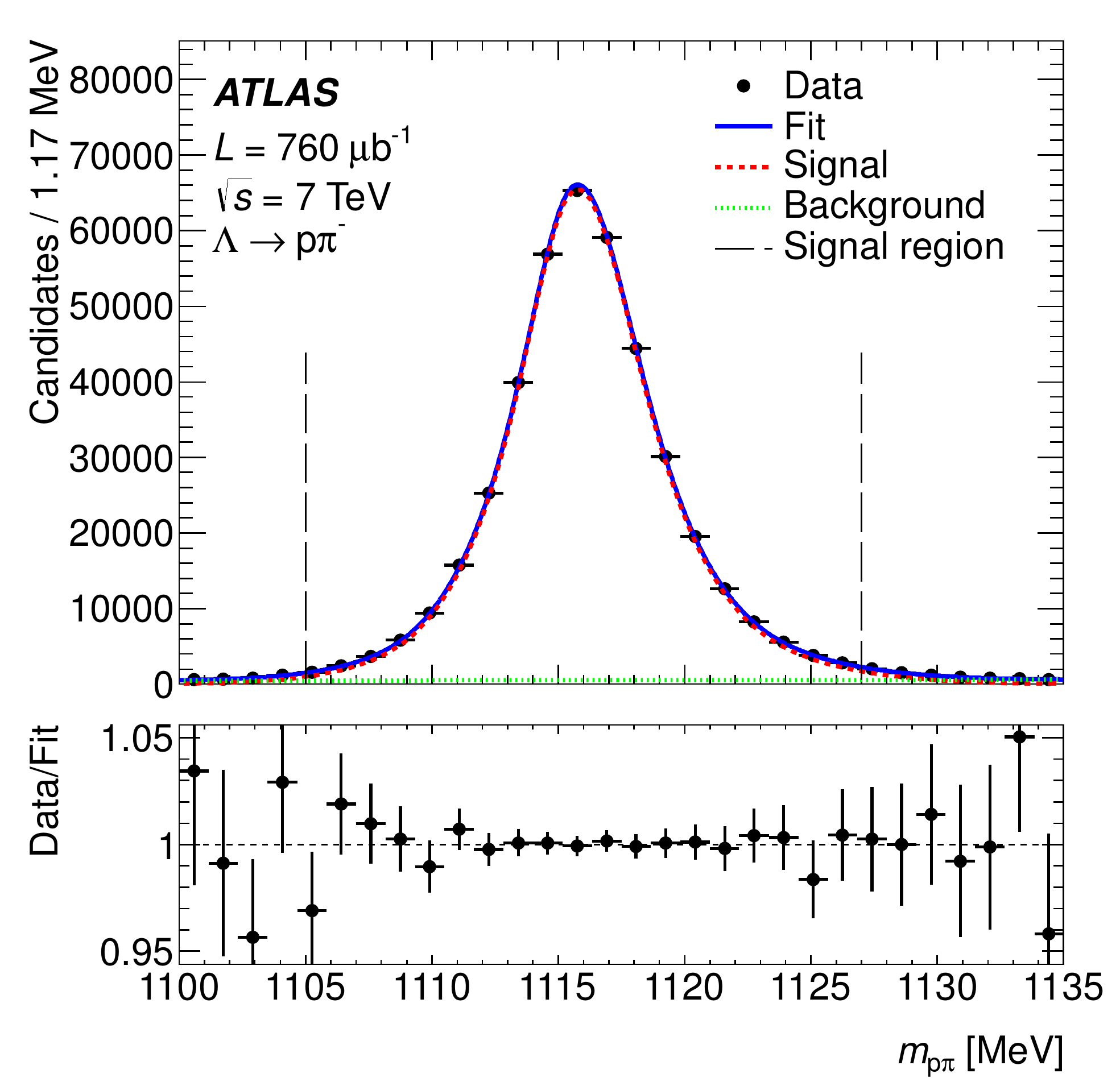}

  \caption{Fit to the invariant mass distribution for \Lam\ candidates used to extract the signal fractions in the signal region and sidebands. The vertical dashed lines mark the boundaries between the mass regions. The fit for \LamBar\ candidates yields similar results, which are listed in Table~\ref{tab:mass}.}
  \label{fig:mass}
\end{figure}

\begin{table*}[tb] 
  \centering
  \caption{Results of the \Lam\ and \LamBar\ invariant mass fits. The table lists the extracted values of mass $m_\Lambda$, slope of the background contribution, $b$, and the signal fraction $f_{\rm sig}$. The mass resolution $\sigma_m$ is calculated numerically from the fit function, $f_{\rm sig}^{\rm gen}$ is the signal fraction determined from the generated particles, and $P_{\chi^2}$ denotes the probability of the fit. The errors are statistical only.}

  \begin{tabular}{@{}lcccccc@{}}
    \hline\hline
    \textbf{Parameter} &\quad& \multicolumn{2}{c}{\bf  \boldmath{\Lam} candidates} &\quad& \multicolumn{2}{c}{\bf  \boldmath{\LamBar} candidates} \\
    \cline{3-4}\cline{6-7}
                       & & \textbf{Data}       & \textbf{MC} & &\textbf{Data}       & \textbf{MC}\\
    \hline
    $m_\Lambda$ [MeV]         & & $1115.75 \pm 0.04$  &  $1115.77 \pm 0.05$  & & $1115.73 \pm 0.03$  &  $1115.79 \pm 0.05$ \\
    $b$ [MeV$^{-1}$]          & & $0.006 \pm 0.002$   &  $0.010 \pm 0.002$   & & $0.009 \pm 0.002$   &  $0.010 \pm 0.002$ \\
    $f_{\rm sig}$             & & $0.962 \pm 0.002$   &  $0.955 \pm 0.002 $  & & $0.964 \pm 0.002$   &  $0.954 \pm 0.002 $\\
    \hline
    $f_{\rm sig}^{\rm gen}$   & &                     &  $0.965 \pm 0.001 $  & &                     &  $0.964 \pm 0.001 $\\
    $\sigma_m$ [MeV]          & & $3.68 \pm 0.02$     &  $3.55  \pm 0.03$    & & $3.75 \pm 0.03$     &  $3.56  \pm 0.03$  \\
    $P_{\chi^2}$              & & $0.79$              &  $0.48$              & & $0.18$              &  $0.51$    \\
    \hline\hline
  \end{tabular}

  \label{tab:mass}
\end{table*}

\begin{figure}[tb] 
  \centering
  \includegraphics[width=\linewidth]{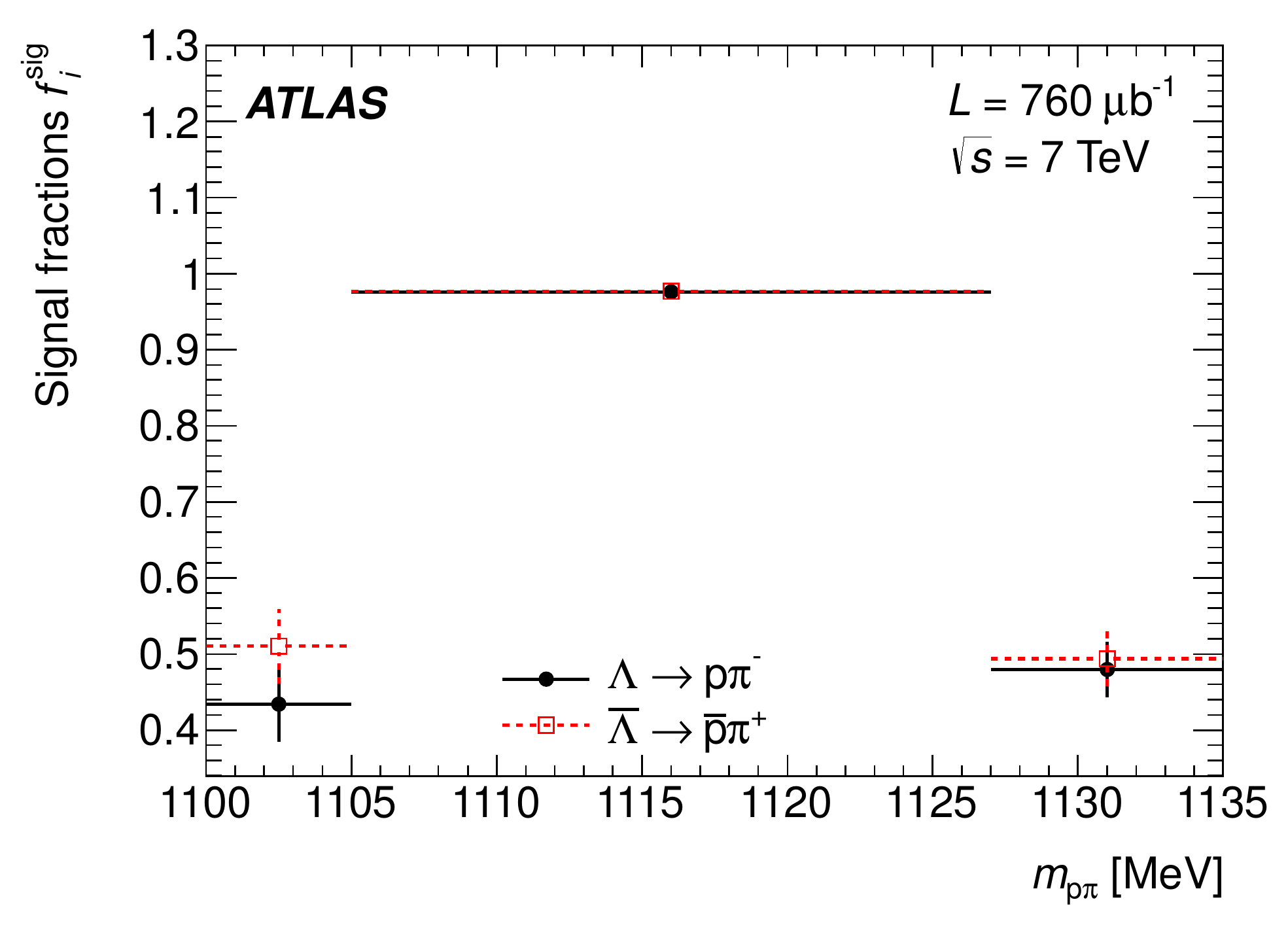}
  \caption{The signal fraction in the left sideband, the signal region, and the right sideband. Displayed statistical uncertainties are obtained from the mass fits.}
  \label{fig:fSig}
\end{figure}

The background component is modeled as a first-order polynomial probability density function:
\begin{equation}
  \mathcal{M}_{\rm bkg}(m_{p\pi}; b) = \frac{1}{\Delta m} \left[ 1 + b (m_{p\pi} - m_c) \right], \notag
\end{equation}
where $m_c$ is the center of the considered mass range, $\Delta m$ its width, and $b$ is the slope of the linear function, which is treated as a free parameter in the fit. Altogether, the probability density function has 11 free parameters. The overall normalization of the fit function is fixed to the area of the histogram. 

The results of the fit are summarized in Table~\ref{tab:mass}. The fit for \Lam\ candidates is shown in Fig.~\ref{fig:mass}. The fit probabilities are $0.79$ for \Lam\ and $0.18$ for \LamBar. The invariant mass resolution $\sigma_m$ is calculated as the square root of the variance of the signal's probability density function. The signal fractions in the signal region and sidebands are calculated as ratios of integrals of the signal component and the full probability density function over the respective mass intervals. The signal fractions are displayed in Fig.~\ref{fig:fSig}. The statistical uncertainties are propagated from uncertainties of the mass fit parameters.

In the MC sample, the value of the signal fraction $f_{\rm sig}^{\rm gen}$ is determined using generated particles and is shown in Table~\ref{tab:mass}. An absolute systematic uncertainty of 0.01 is assigned to the signal fraction to account for the difference between $f_{\rm sig}^{\rm gen}$ and $f_{\rm sig}$ in the simulated sample. Since tails of the mass distribution are hard to model, this systematic uncertainty is larger in the sidebands. It is 0.12 for the left sideband and 0.25 for the right sideband. Other sources of systematic uncertainty are discussed in Sec.~\ref{sec:systematics}. 

\section{Polarization extraction}
\label{sec:polarization}

\begin{figure*}[tb] 
  \centering
  \includegraphics[width=.497\linewidth]{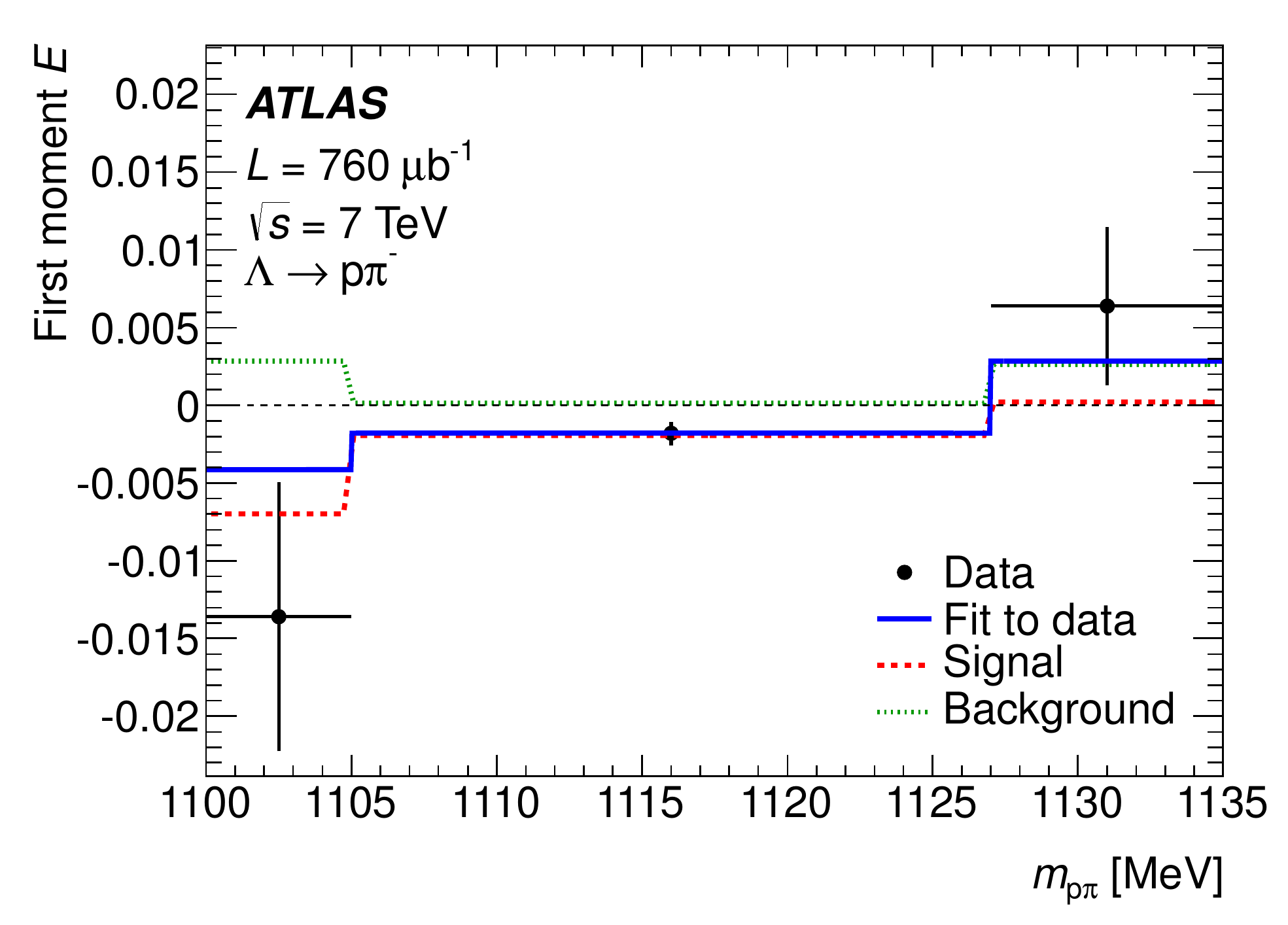}
  \includegraphics[width=.497\linewidth]{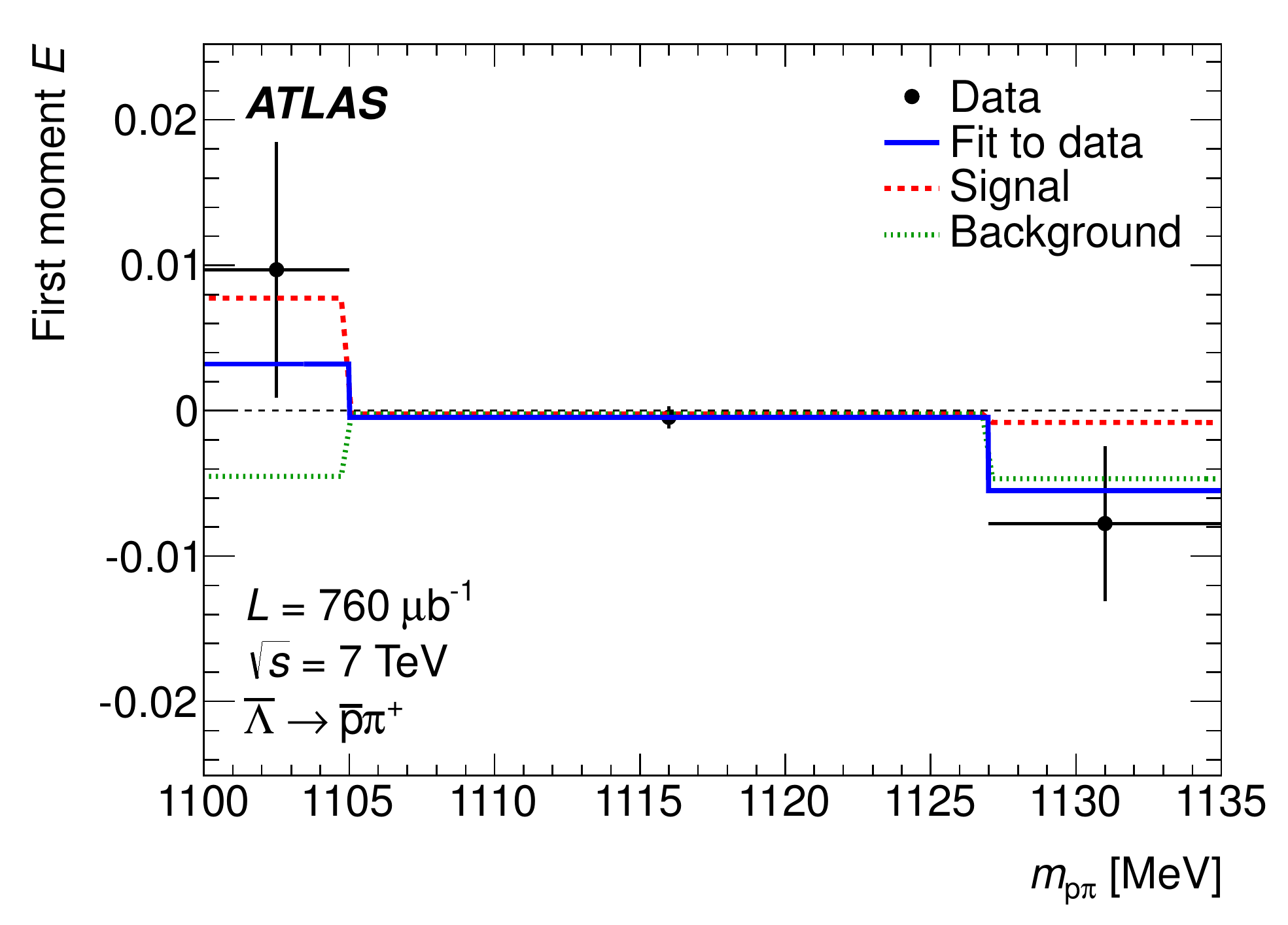}

  (a)\hspace{0.457\linewidth}(b)

  \caption{The first moment of the (a) \Lam\ and (b) \LamBar\ decay angle distribution, $E$, and the corresponding least-squares fit. The signal component corresponds to the product $f^{\rm sig}_i \big\{ E^{\rm MC}_i(0)  + \big[E^{\rm MC}_i(1) - \ E^{\rm MC}_i(0)\big]P \big\}$, the background component to $(1-f^{\rm sig}_i) E_{\rm bkg}$, and the fit is the sum of the two [see Eq.~(\ref{eqn:E_final})].}
  \label{fig:pol_fit}
\end{figure*}

\begin{figure}[tb] 
  \centering
  \includegraphics[width=\linewidth]{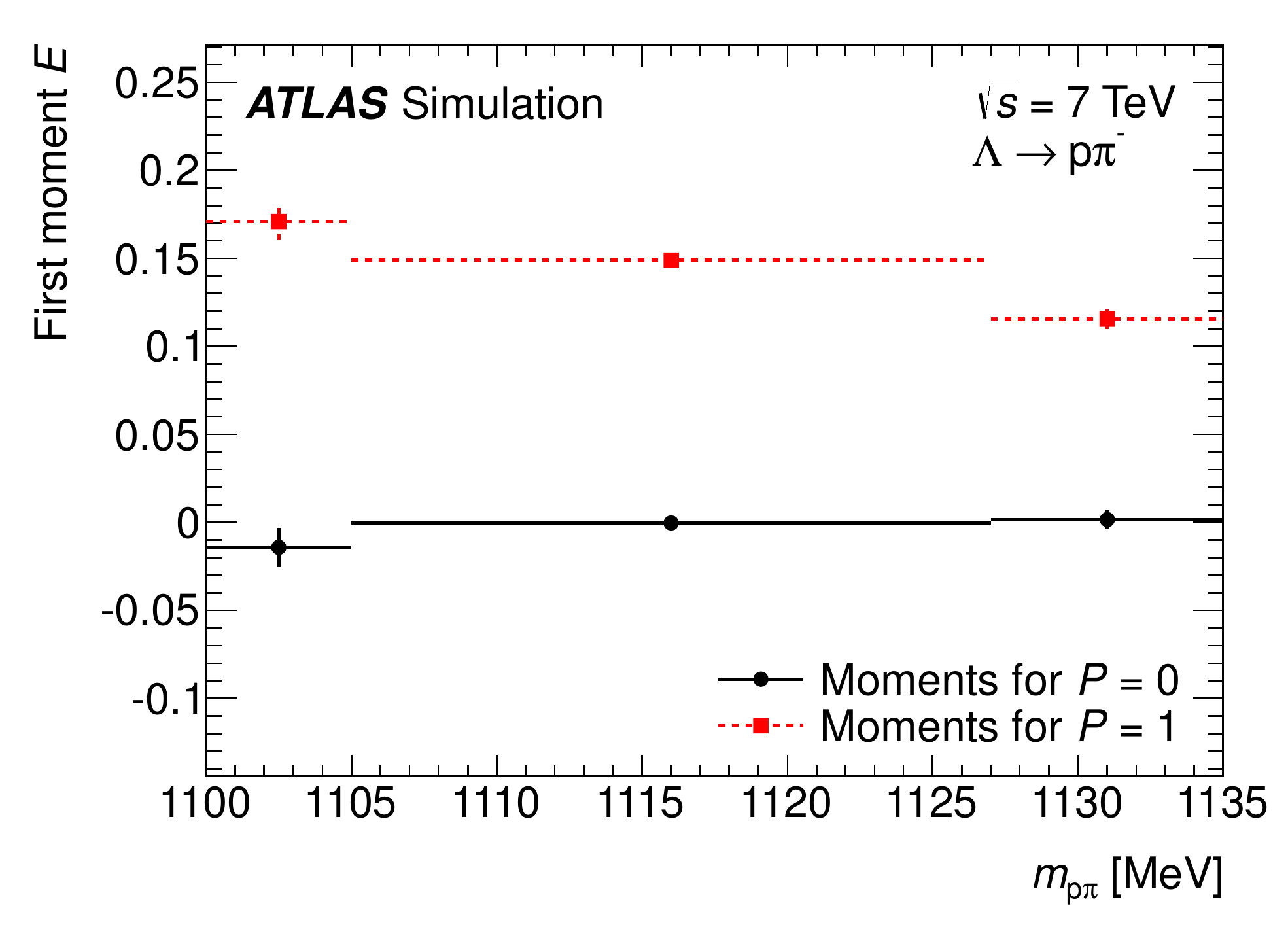}
  \caption{First moments of the angular distributions of MC signal events with known polarization.}
  \label{fig:mc_moments}
\end{figure}

\subsection{Least-squares fit}

A closure test for the polarization extraction procedure is conducted with the simulated sample. The first half of the MC sample is used as a test sample and is weighted to a transverse polarization of $-0.3$. The second half is used to calculate the MC moments. The polarization is then extracted using the method of moments. The extracted polarization, $P_{\rm test} = -0.302 \pm 0.005({\rm stat})$, is in agreement with the input value. 

Polarization for \Lam\ and \LamBar\ is extracted via the least square fit of the first moment of the decay angle distribution measured in the signal region and sidebands, shown in Fig.~\ref{fig:pol_fit}. The figure contains values for the data, simulations for signal and background, and results of the fit. The fit function, Eq.~(\ref{eqn:E_final}), has two free parameters: the polarization $P$ and the first moment of the angular distribution of the background, $E_{\rm bkg}$. The other parameters in Eq.~(\ref{eqn:E_final}), the signal fractions $f^{\rm sig}_i$ and the first moments $E^{\rm MC}_i(0)$ and $E^{\rm MC}_i(1)$, are fixed. The values of $f^{\rm sig}_i$ are extracted from the distributions in $m_{p\pi}$ as described in Sec.~\ref{sec:mass_fit}. The moments $E^{\rm MC}_i(0)$ and $E^{\rm MC}_i(1)$ are calculated using the MC simulations for decays of hyperons with polarization of 0 and 1, respectively. The values of $E^{\rm MC}_i(0)$ and $E^{\rm MC}_i(1)$ for \Lam\ are shown in Fig.~\ref{fig:mc_moments}.

The extracted values of polarization are $P = -0.010 \pm 0.005$(stat) for \Lam\ and $P = 0.002 \pm 0.006$(stat) for \LamBar. The first moments of the angular distribution of backgrounds are $E_{\rm bkg} = 0.005 \pm 0.008$(stat) for the \Lam\ sample and $E_{\rm bkg} = -0.009 \pm 0.009$(stat) for the \LamBar\ sample. These values were obtained  by minimizing the $\chi^2$ function, Eq.~(\ref{eqn:chi2}). The correlation coefficient between $P$ and $E_{\rm bkg}$ is estimated to be 0.33 for both the \Lam\ and \LamBar\ fits. The cumulative probabilities corresponding to the minimal $\chi^2$ values from these fits are 0.20 and 0.39 for \Lam\ and \LamBar, respectively.

Figure~\ref{fig:pol_fit} shows statistical uncertainties of the measured first moments only. Statistical uncertainties of the first moments of simulated \Lam\ events, which affect the fitted function, are not shown in Fig.~\ref{fig:pol_fit}. The impact of these uncertainties on the final result is evaluated as a systematic uncertainty in Sec.~\ref{sec:systematics}.

\begin{figure}[h!] 
  \centering
  \includegraphics[width=\linewidth]{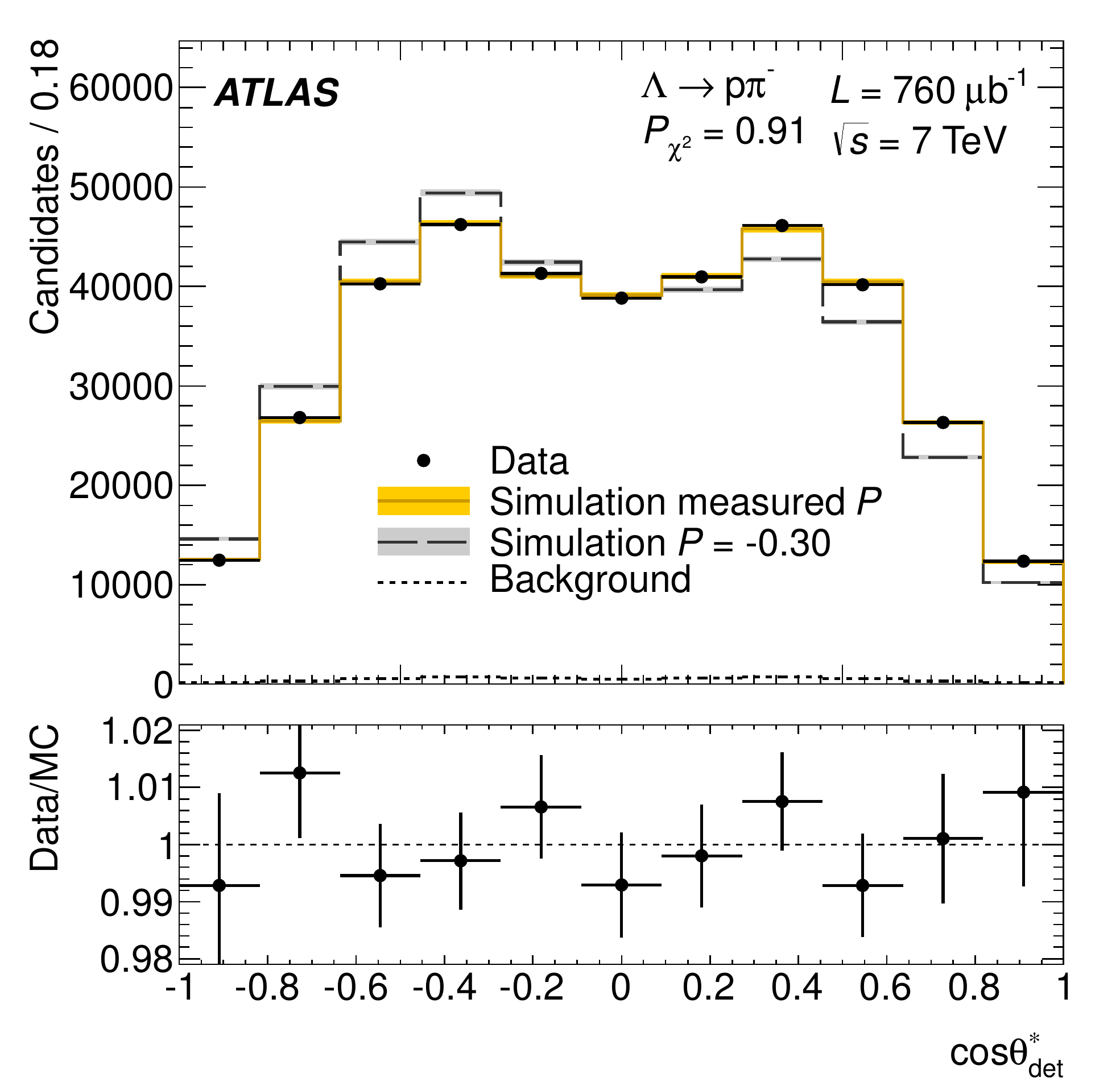}
  
  \caption{Comparison of the measured decay angle distribution in data and MC samples for \Lam\ events in the signal region. The MC events are reweighted using the extracted value of the polarization. The background contribution is estimated from the sideband regions. The probability of the $\chi^2$-test, $P_{\chi^2}$, is 0.91. For comparison, the simulated distribution for $P=-0.30$ is also shown.}
  \label{fig:comparison}
\end{figure}

The MC samples are reweighted using Eq.~(\ref{eqn:wPol}) with the extracted values of the polarization, as shown in Fig.~\ref{fig:comparison} for \Lam\ hyperons. Using the $\chi^2$-test, the reweighted decay angle distributions are found to agree with the data. 

The polarization is also extracted from samples binned in \pt\ and \xF. The results for the binned samples are summarized in Sec.~\ref{sec:results}. 

\subsection{Systematic uncertainties}
\label{sec:systematics}

\begin{table}[tb] 
  \centering
  \caption{Summary of systematic uncertainties. The numbers represent absolute systematic uncertainties of the polarization values. All negligible systematic uncertainties are summarized under ``Other contributions''. Individual values before rounding are added in quadrature to obtain the total systematic uncertainty.}

  \begin{tabular}{@{}lccc@{}}
  \hline\hline
  \bf Systematic uncertainty &\quad& \boldmath{\Lam} & \boldmath{\LamBar}\\
  \hline
  MC statistics           	      && 0.003      	& 0.003      		\\
  Mass range               	      && 0.003      	& 0.003      		\\
  Background              	      && 0.001      	& 0.001      		\\
  Kinematic weighting     	      && 0.001        & 0.001		\\
  Other contributions             && $<5 \times 10^{-4}$ 	& $<5 \times 10^{-4}$ 	\\
  \hline
  Total                   	      && 0.004        & 0.004     \\
  \hline\hline

  \end{tabular}

  \label{tab:systematics}
\end{table}

Systematic uncertainties are estimated by modifying various aspects of the analysis and observing how they affect the extracted value of the polarization. The estimated values are summarized in Table~\ref{tab:systematics}. The listed systematic uncertainties are assumed to be uncorrelated. Individual values before rounding are added in quadrature to obtain the total systematic uncertainty. Details of the individual uncertainties follow.

\begin{enumerate}
\item The MC sample size contributes to the uncertainty of the first moments of the simulated angular distributions. To estimate what effect the MC statistics have on the polarization measurement, ten pseudoexperiments are performed with values of $E^{\rm MC}_i(0)$ and $E^{\rm MC}_i(1)$ varied according to the normal distribution whose width corresponds to the estimated statistical uncertainties of the  moments. The uncertainty on the extracted polarization value is estimated as the standard deviation of the results of these pseudoexperiments.

\item The signal region size is varied up and down by 2~MeV to estimate the impact of the choice of width of the mass window on the polarization measurement. The alternative signal regions are chosen such that enough events are retained in the sidebands. The uncertainty is taken as the difference between the alternative and default results.

\begin{table*}[tb] 
  \centering
  \caption{Transverse polarization of \Lam\ and \LamBar\ measured in the full fiducial phase space and in bins of \xF\ and \pt. The values of $\bar x_{\rm F}$ and $\bar p_{\rm T}$ are mean values of \xF\ and \pT, respectively, in given ranges. The table lists both the statistical and systematic uncertainties.}
  
  \begin{tabular}{@{}lcccccc@{}}
    \hline\hline
    \bf Sample & \boldmath{$\bar x_{\rm F}$} & & \bf\boldmath{$\bar p_{\rm T}$}& & \multicolumn{2}{c}{\bf Polarization} \\
    \cline{2-2}
    \cline{4-4}
    \cline{6-7}
     \bf         &  \boldmath{[$10^{-4}$]} && \bf [GeV] & &\boldmath{\Lam} & \boldmath{ \LamBar} \\
     \hline
     
     Full fiducial volume                  & $10.0$ && $1.91$ & &    $-0.010 \pm 0.005 \pm 0.004$ & \sp$ 0.002 \pm 0.006 \pm 0.004$ \\
    \hline
     
    $x_{\rm F} \in (0.5, 5) \times 10^{-4}$    & $2.8$  && $1.83$ & & \sp$ 0.005 \pm 0.009 \pm 0.006$ &    $-0.009 \pm 0.010 \pm 0.006$ \\
    $x_{\rm F} \in (5, 10.5) \times 10^{-4}$   & $7.5$  && $1.85$ & &    $-0.012 \pm 0.009 \pm 0.008$ & \sp$ 0.002 \pm 0.010 \pm 0.007$ \\
    $x_{\rm F} \in (10.5, 100) \times 10^{-4}$ & $19.3$ && $2.12$ & &    $-0.005 \pm 0.010 \pm 0.008$ & \sp$ 0.012 \pm 0.010 \pm 0.010$ \\
    \hline

    $p_{\rm T} \in (0.8, 1.3)$~GeV             & $ 7.5$ && $1.07$ & &    $-0.008 \pm 0.012 \pm 0.011$ &    $-0.004 \pm 0.013 \pm 0.013$  \\
    $p_{\rm T} \in (1.3, 2.03)$~GeV            & $ 9.3$ && $1.64$ & &    $-0.019 \pm 0.009 \pm 0.007$ &    $-0.003 \pm 0.010 \pm 0.007$  \\
    $p_{\rm T} \in (2.03, 15)$~GeV             & $12.6$ && $2.84$ & &    $-0.005 \pm 0.008 \pm 0.005$ & \sp$ 0.009 \pm 0.009 \pm 0.004$  \\
    \hline\hline
  \end{tabular}

  \label{tab:results}
\end{table*}

\item The systematic uncertainty due to the background consists of three components which are added in quadrature:

An alternative linear background model, $E_{\rm bkg}(m_{p\pi}) = E_{\rm bkg}^0 + E_{\rm bkg}^1 m_{p\pi}$ with $E_{\rm bkg}^0$ and $E_{\rm bkg}^1$ being the fit parameters, is used to probe the dependence of the final result on the background parameterization.

Another source of background systematic uncertainty is the statistical uncertainty of the signal fraction $f_{\rm sig}$ and the slope of the background contribution, $b$, determined in the mass fit in Sec.~\ref{sec:mass_fit}. These uncertainties directly affect the calculated signal fractions $f^{\rm sig}_i$, which in turn affect the polarization fit. The parameters are varied to propagate these uncertainties to the final result.

The last of the background uncertainties is the systematic uncertainty of the signal fractions, estimated to be 0.01, 0.12, and 0.25 in the signal region, left sideband and right sideband, respectively. These uncertainties are propagated into the final result by varying the signal fractions in the polarization fit.

\item The uncertainty associated with the MC weighting is estimated. The MC sample is weighted so that the \Lam\ momentum distribution agrees with data after the background subtraction. If the background subtraction is not perfect, this affects the weighted signal spectrum. To assess the dependence of the weight function on the background estimation, an alternative weight function is constructed using data without background subtraction and the measurement is repeated. 
\end{enumerate}

In addition to these systematic uncertainties, numerous other effects were studied, but were found to have negligible impact on the extracted polarization: 

\begin{enumerate}
\setcounter{enumi}{4}
\item Uncertainties on the track momentum scale and resolution are estimated using the fits to the \Lam\ invariant mass.

\item To estimate the uncertainty on the track reconstruction efficiency in the simulation, the MC and data samples are binned in \pT, $\eta$, and the number of silicon hits on proton and pion tracks. Relative differences between the number of tracks in bins of the data and MC distributions are then attributed to the tracking efficiency uncertainty and propagated to the final result using event weighting.

\item Equation~(\ref{eqn:E_final}) assumes that the total efficiency of \Lam\ reconstruction does not depend on polarization. This is satisfied if the efficiency $\varepsilon(t)$ is an even function of $t$. It was checked that this assumption has a negligible effect on the result.

\item The uncertainty of the $\alpha$ parameter measured previously \cite{PDG, Astbury:1975hn, *Cleland:1972fa, *Dauber:1969hg, *Overseth:1967zz, *Cronin:1963zb} is propagated to the uncertainty of the measured polarization.

\item The measurement is performed in the phase space defined by the \pT, $\eta$, and \xF\ ranges. With the data, these can only be defined in terms of measured quantities that are affected by the finite detector resolution. The fraction of events that migrate into and out of the fiducial volume due to resolution is estimated using simulation and the impact of event migration on the polarization is estimated.

\item The small fraction ($<3$\%) of \Lam\ hyperons that are produced from weak decays can be polarized in the direction parallel to their momenta, whereas this measurement is performed assuming the parallel component of the polarization vector is zero. Using the simulation, it is estimated that weakly produced \Lam\ hyperons cannot significantly affect the final result. 

Similarly, the contribution of \Lam\ hyperons from secondary hadronic interactions with the detector material ($<3$\%), which have zero transverse polarization in the chosen reference frame, can be neglected.

\begin{figure*}[tb] 
  \centering
  \includegraphics[width=.48\linewidth]{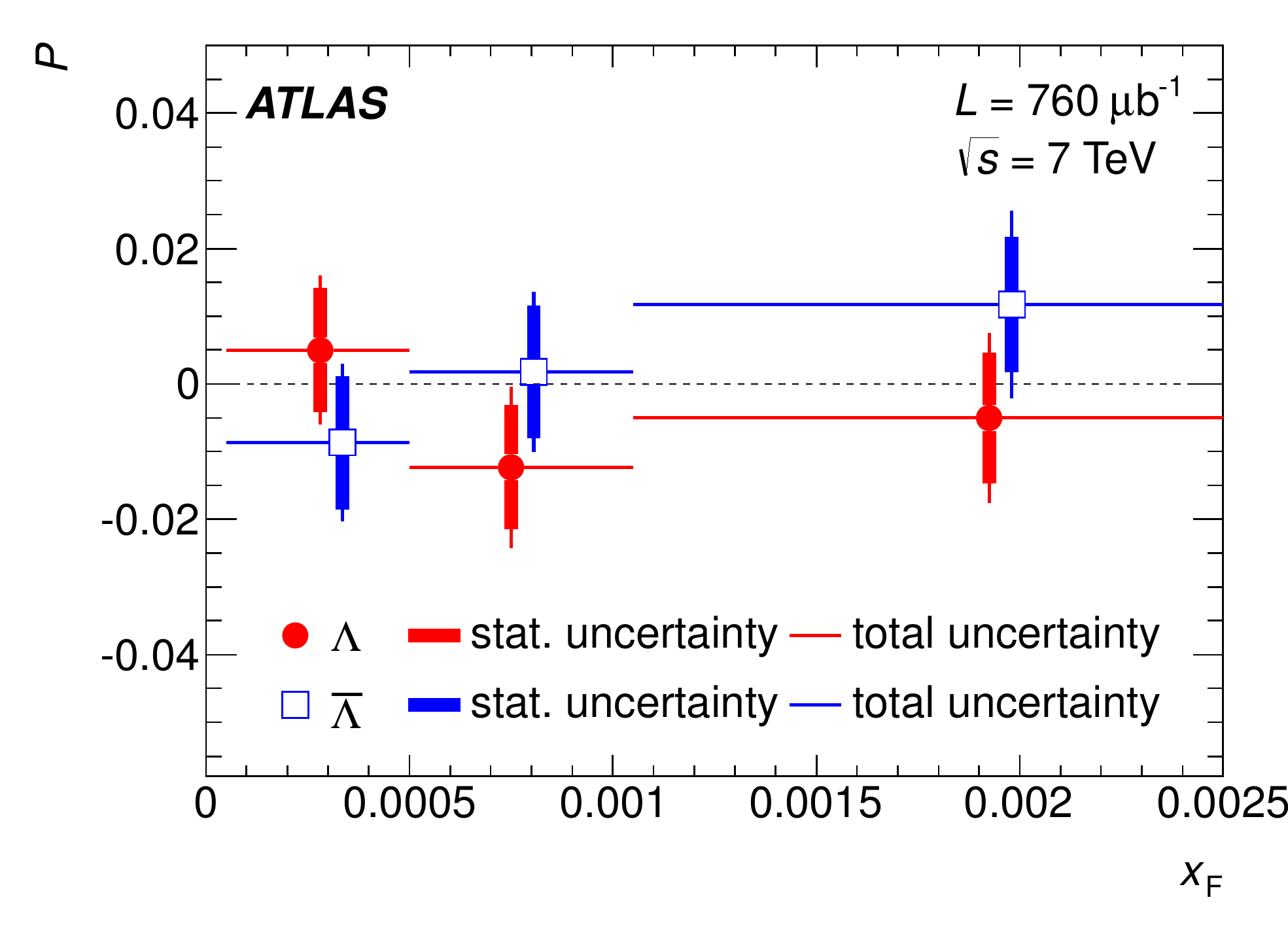}
  \includegraphics[width=.48\linewidth]{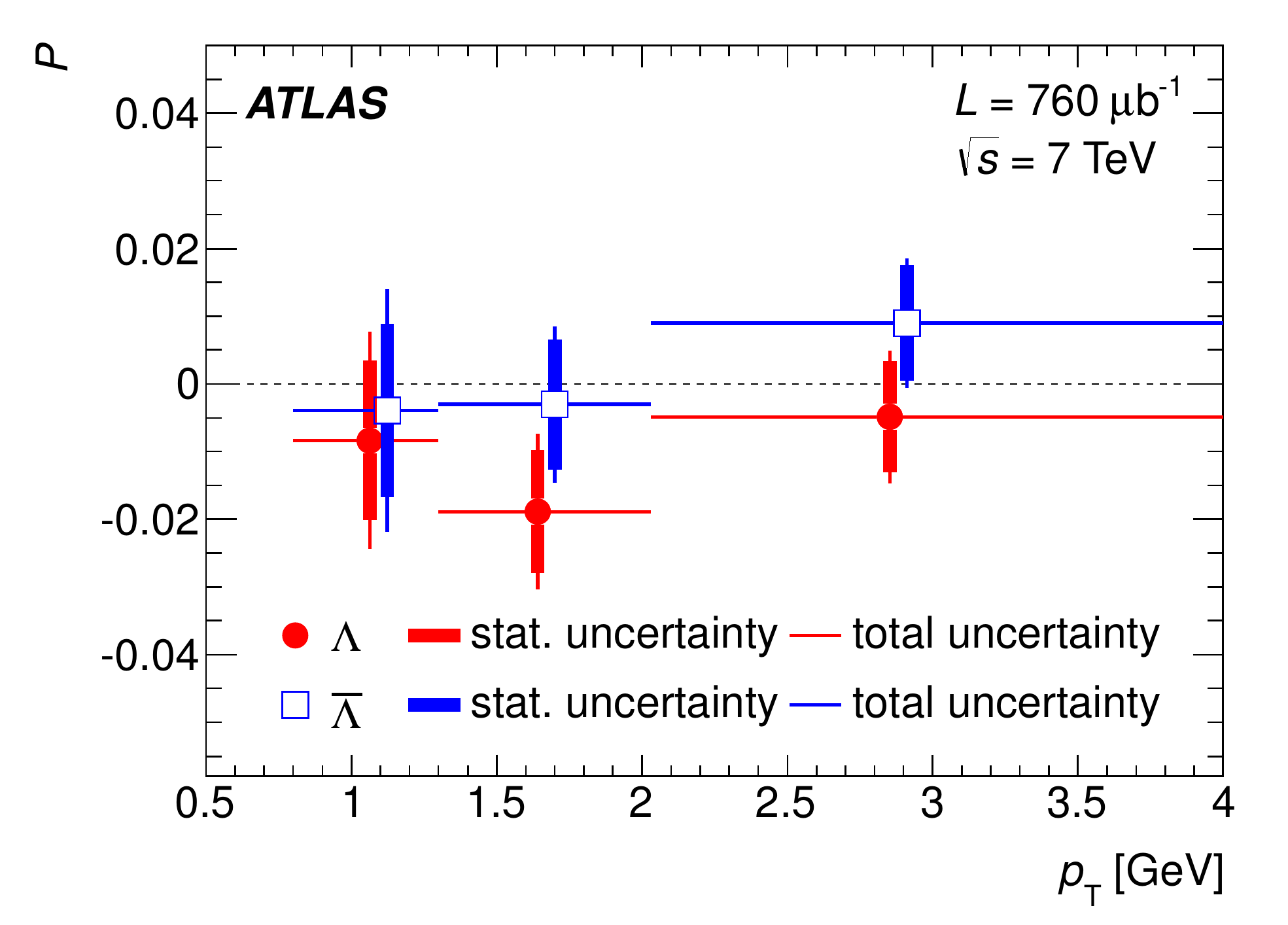}

  (a)\hspace{0.457\linewidth}(b)

  \caption{Transverse polarization of \Lam\ and \LamBar\ hyperons as a function of (a) $x_{\rm F}$ and (b) $p_{\rm T}$. The thick error bars represent statistical uncertainty of the measurements, the narrow error bars are total uncertainties. Data points for \LamBar\ are displaced horizontally for better legibility of the results.}
  \label{fig:results}
\end{figure*}

\begin{figure}[tb] 
  \centering
  \includegraphics[width=\linewidth]{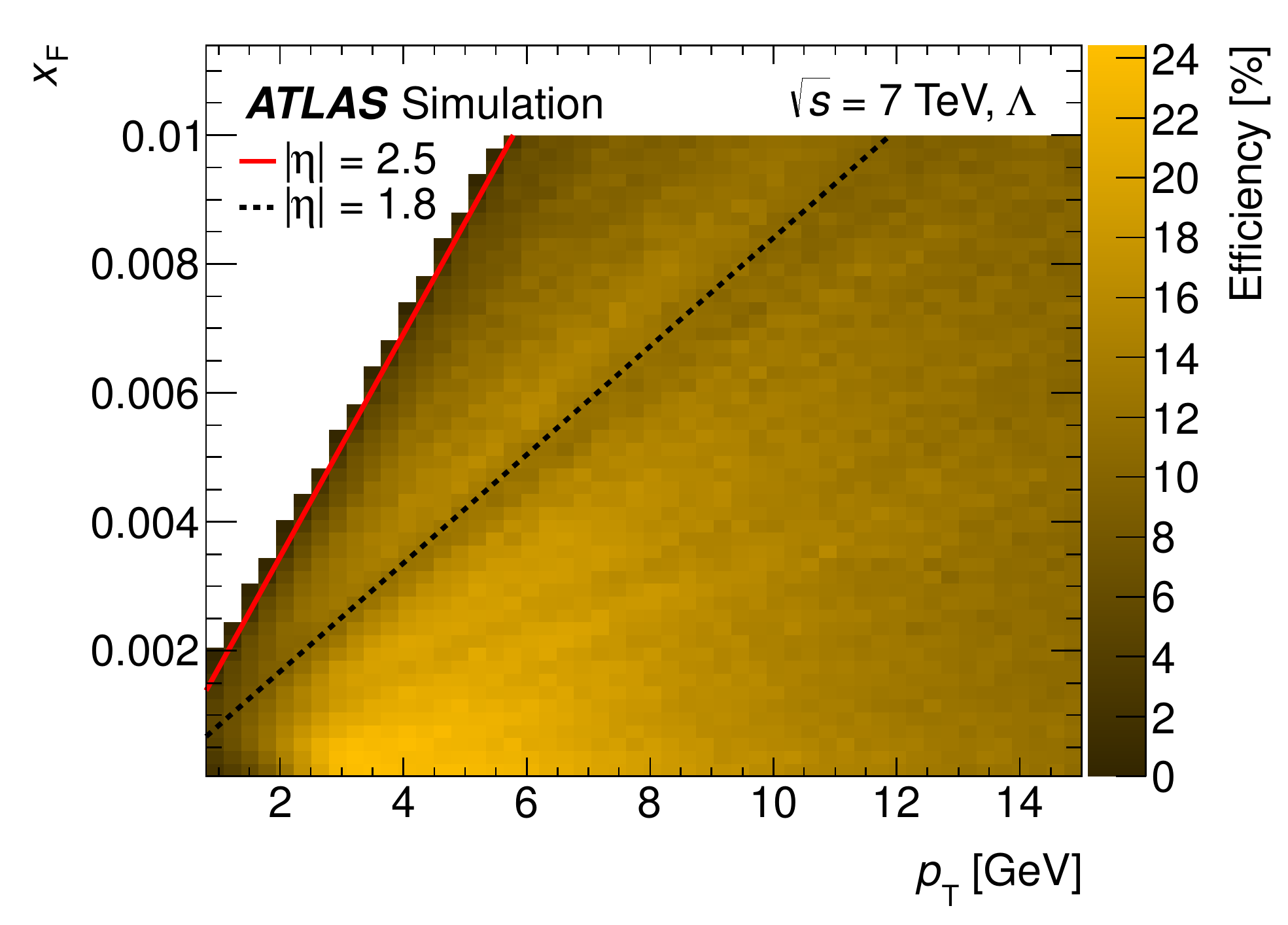}

  \caption{The \Lam\ reconstruction and selection efficiency as a function of $p_{\rm T}$ and $x_{\rm F}$. A drop in efficiency for events with $|\eta| \approx 1.8$ is caused by the geometry of the ID. Absolute statistical uncertainties are less than 0.6\% in all bins.}
  \label{fig:phasespace_lam}
\end{figure}

\item Lastly, the MC simulation is used to show that the trigger selection does not bias the polarization measurement.
\end{enumerate}


\section{Results}
\label{sec:results}

The average transverse polarizations of \Lam\ and \LamBar\ are measured to be
\begin{eqnarray}
  P_\Lambda       &=&    -0.010 \pm 0.005({\rm stat}) \pm 0.004({\rm syst})\ {\rm and} \notag\\
  P_{\bar\Lambda} &=& \sp 0.002 \pm 0.006({\rm stat}) \pm 0.004({\rm syst}) \notag
\end{eqnarray}
in the fiducial phase space defined by the ranges $0.8< p_{\rm T} < 15$~GeV, $5\times 10^{-5} < x_{\rm F} < 0.01$, and $|\eta| < 2.5$. The polarization is also measured in three bins with mean \pT\ of $1.07$, $1.64$, and $2.85$~GeV and in three bins with mean $x_{\rm F}$ of $2.8 \times 10^{-4}$, $7.5 \times 10^{-4}$, and $19.3 \times 10^{-4}$. The results are shown in Fig.~\ref{fig:results} and listed in Table~\ref{tab:results}. The systematic uncertainties due to MC statistics are anti-correlated between the \Lam\ and \LamBar\ results since they are estimated using the same MC sample. 

The measured polarization values reported here depend on the reconstruction efficiency within the fiducial phase space, $\epsilon(x_{\rm F}, p_{\rm T})$, and on the differential polarization $P(x_{\rm F}, p_{\rm T})$ since the average polarization is defined~as
\begin{eqnarray}
  P = \frac{1}{\mathcal{N}} \int dx_{\rm F}\,dp_{\rm T}\, g(x_{\rm F}, p_{\rm T}) \epsilon(x_{\rm F}, p_{\rm T}) P(x_{\rm F}, p_{\rm T}),
  \label{eqn:average_P}
\end{eqnarray}
where $g(x_{\rm F}, p_{\rm T})$ is the probability density function of \xF\ and \pT\ and $\mathcal{N} = \int dx_{\rm F}\,dp_{\rm T}\, g(x_{\rm F}, p_{\rm T}) \epsilon(x_{\rm F}, p_{\rm T})$. To allow comparisons between the measured values and any theoretical parameterization of $P(x_{\rm F}, p_{\rm T})$, the efficiency maps of reconstructed \Lam\ and \LamBar\ decays are provided in the HEPDATA database \cite{hepdata}. Figure~\ref{fig:phasespace_lam} shows the reconstruction efficiency for \Lam. The MC simulation is used to check that the reconstruction efficiency as a function of \pt\ and \xF\ does not depend on the value of the polarization. These maps can therefore be used to calculate the expected average polarization given by Eq.~(\ref{eqn:average_P}) for any theoretical model.

\begin{figure}[tb] 
  \centering
  \includegraphics[width=\linewidth]{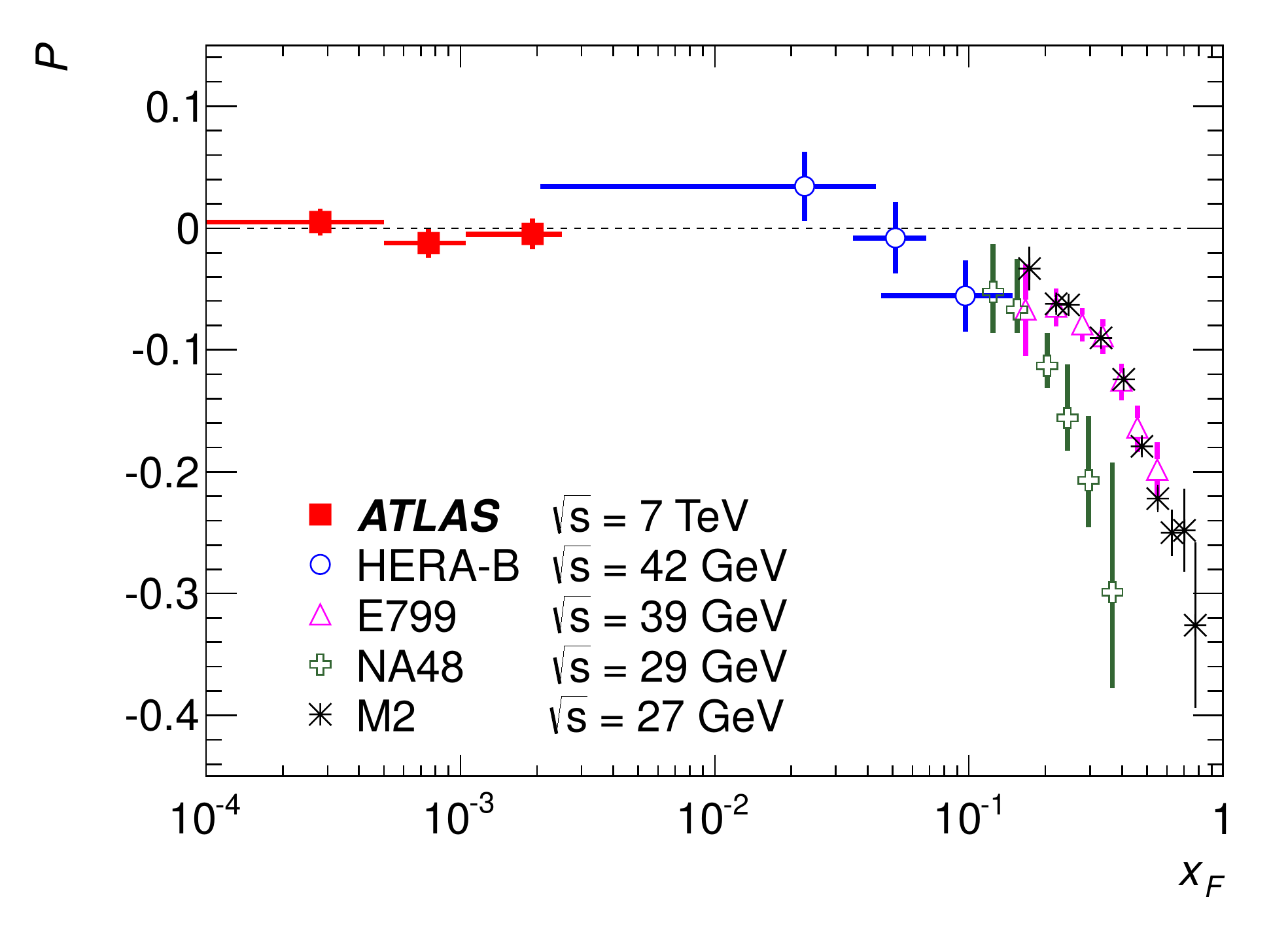}

  \caption{The \Lam\ transverse polarization measured by ATLAS compared to measurements from lower center-of-mass energy experiments. HERA-B data are taken from Ref.~\cite{Hera-B}, NA48 from Ref.~\cite{Fan99}, E799 data from Ref.~\cite{Ram94}, and M2 from Ref.~\cite{Lun89}. The HERA-B results are transformed to positive values of \xF\ using Eq.~(\ref{eqn:pol_odd}).}
  \label{fig:measurements}
\end{figure}

Figure~\ref{fig:measurements} compares this result with other measurements: an experiment at the M2 beam-line at Fermilab \cite{Lun89}, experiment E799 \cite{Ram94} also at Fermilab, NA48 \cite{Fan99} at CERN, and HERA-B \cite{Hera-B} at DESY. Direct comparison of results from different experiments is non-trivial, since each measurement was made at a different center-of-mass energy and covers a different phase space: the average \pT\ coverage of the M2 experiment is 0.62--1.74~GeV, the \pT\ range of the E799 experiment is 0.67--2.15~GeV, NA48 covers \pT\ of 0.28--0.86~GeV,  HERA-B 0.82--0.84~GeV, and in Fig.~\ref{fig:measurements} the ATLAS data cover the average \pT\ range of 1.83--2.12~GeV. Furthermore, the HERA-B measurement covers negative values of \xF. In Fig.~\ref{fig:measurements}, the HERA-B results are transformed using Eq.~(\ref{eqn:pol_odd}) so that they can be compared with other results. The E799 and NA48 experiments define \xF\ as the fraction of the beam energy carried by the \Lam. In Fig.~\ref{fig:measurements}, the values are transformed according to the definition of \xF\ used in this paper (although the difference is very small). Figure~\ref{fig:measurements} shows only a subset of the M2 results measured at a fixed production angle of about 6~mrad with a beryllium target. A parameterization of the polarization \cite{Lun89} as a function of \xF\ and \pT\ is used to extrapolate the results of the M2 experiment to the \xF\ and \pT\ range of this measurement. The extrapolated value is less than $5\times 10^{-4}$, which can effectively be treated as zero given the precision of this measurement.

The fixed-target experiments \cite{Bun76, *Hel78, *Sch78, *Hel83, *Yul91, Lun89, Ram94, Fan99, Hera-B} did not observe any strong dependence of the \Lam\ polarization on the collision energy. Some energy dependence could be introduced at the LHC, since due to a significant collision energy increase, the fraction of \Lam\ hyperons from decays of heavier baryons may be different than for low energy collisions. Using {\sc Pythia}, it is estimated that about $50$\% of the \Lam\ hyperons in ATLAS data are produced in decays, which is comparable to the estimate of about $40$\% for NA48 \cite{Fan99}. Therefore, assuming that the polarization of the original baryons is diluted in the decay, the magnitude of the polarization at the LHC is expected to be slightly smaller than that observed by the previous experiments at the same \pt\ and \xF. In the absence of any new polarization-producing mechanism that would manifest itself at low \xF\ and high center-of-mass energies, the measured polarization is expected to be consistent with zero; this is so for the results presented here.

\section{Conclusions}

Results of the measurement of the transverse polarization of \Lam\ and \LamBar\ hyperons carried out with the ATLAS experiment at the LHC in 760~$\mu$b$^{-1}$ of proton--proton collisions at $\sqrt{s} = 7$~TeV are presented here. The polarization is measured for inclusively produced \Lam\ hyperons in the fiducial phase space and in three bins of the transverse momentum with mean \pT\ of $1.07$, $1.64$, and $2.85$~GeV and in three bins of the Feynman-$x$ variable with mean \xF\ of $2.8 \times 10^{-4}$, $7.5 \times 10^{-4}$, and $19.3 \times 10^{-4}$. Average polarizations in the full fiducial volume are measured to be $-0.010 \pm 0.005({\rm stat}) \pm 0.004({\rm syst})$ for \Lam\ and $0.002 \pm 0.006({\rm stat}) \pm 0.004({\rm syst})$ for \LamBar\ hyperons. In \pT\ and \xF\ bins, the polarization is found to be less than 2\% and is consistent with zero within the estimated uncertainties. 

The result for the \Lam\ polarization is consistent with an extrapolation the the results of the M2 beam-line experiment at Fermilab \cite{Lun89} to low $x_{\rm F}$, which suggests that the magnitude of the polarization should decrease as $x_{\rm F}$ approaches zero.

Unlike for the \Lam, the polarization of the \LamBar\ hyperons was measured to be consistent with zero by all the previous experiments. The ATLAS experiment also measured the \LamBar\ polarization to be consistent with zero in the fiducial phase space of the analysis, as well as in bins of \pT\ and \xF.

%
\section*{Acknowledgments}

We thank CERN for the very successful operation of the LHC, as well as the
support staff from our institutions without whom ATLAS could not be
operated efficiently.

We acknowledge the support of ANPCyT, Argentina; YerPhI, Armenia; ARC,
Australia; BMWFW and FWF, Austria; ANAS, Azerbaijan; SSTC, Belarus; CNPq and FAPESP,
Brazil; NSERC, NRC and CFI, Canada; CERN; CONICYT, Chile; CAS, MOST and NSFC,
China; COLCIENCIAS, Colombia; MSMT CR, MPO CR and VSC CR, Czech Republic;
DNRF, DNSRC and Lundbeck Foundation, Denmark; EPLANET, ERC and NSRF, European Union;
IN2P3-CNRS, CEA-DSM/IRFU, France; GNSF, Georgia; BMBF, DFG, HGF, MPG and AvH
Foundation, Germany; GSRT and NSRF, Greece; ISF, MINERVA, GIF, I-CORE and Benoziyo Center,
Israel; INFN, Italy; MEXT and JSPS, Japan; CNRST, Morocco; FOM and NWO,
Netherlands; BRF and RCN, Norway; MNiSW and NCN, Poland; GRICES and FCT, Portugal; MNE/IFA, Romania; MES of Russia and ROSATOM, Russian Federation; JINR; MSTD,
Serbia; MSSR, Slovakia; ARRS and MIZ\v{S}, Slovenia; DST/NRF, South Africa;
MINECO, Spain; SRC and Wallenberg Foundation, Sweden; SER, SNSF and Cantons of
Bern and Geneva, Switzerland; NSC, Taiwan; TAEK, Turkey; STFC, the Royal
Society and Leverhulme Trust, United Kingdom; DOE and NSF, United States of
America.

The crucial computing support from all WLCG partners is acknowledged
gratefully, in particular from CERN and the ATLAS Tier-1 facilities at
TRIUMF (Canada), NDGF (Denmark, Norway, Sweden), CC-IN2P3 (France),
KIT/GridKA (Germany), INFN-CNAF (Italy), NL-T1 (Netherlands), PIC (Spain),
ASGC (Taiwan), RAL (UK) and BNL (USA) and in the Tier-2 facilities
worldwide.


%

\onecolumngrid

\clearpage
\begin{flushleft}
{\Large The ATLAS Collaboration}

\bigskip

G.~Aad$^{\rm 85}$,
B.~Abbott$^{\rm 113}$,
J.~Abdallah$^{\rm 152}$,
S.~Abdel~Khalek$^{\rm 117}$,
O.~Abdinov$^{\rm 11}$,
R.~Aben$^{\rm 107}$,
B.~Abi$^{\rm 114}$,
M.~Abolins$^{\rm 90}$,
O.S.~AbouZeid$^{\rm 159}$,
H.~Abramowicz$^{\rm 154}$
% ...
\bigskip
\\
% ...
$^{11}$ Institute of Physics, Azerbaijan Academy of Sciences, Baku, Azerbaijan\\
% ...
$^{85}$ CPPM, Aix-Marseille Universit{\'e} and CNRS/IN2P3, Marseille, France\\
% ...
$^{90}$ Department of Physics and Astronomy, Michigan State University, East Lansing MI, United States of America\\
% ...
$^{107}$ Nikhef National Institute for Subatomic Physics and University of Amsterdam, Amsterdam, Netherlands\\
% ...
$^{113}$ Homer L. Dodge Department of Physics and Astronomy, University of Oklahoma, Norman OK, United States of America\\
$^{114}$ Department of Physics, Oklahoma State University, Stillwater OK, United States of America\\
% ...
$^{117}$ LAL, Universit{\'e} Paris-Sud and CNRS/IN2P3, Orsay, France\\
% ...
$^{152}$ Institute of Physics, Academia Sinica, Taipei, Taiwan\\
% ...
$^{154}$ Raymond and Beverly Sackler School of Physics and Astronomy, Tel Aviv University, Tel Aviv, Israel\\
% ...
$^{159}$ Department of Physics, University of Toronto, Toronto ON, Canada\\
% ...
\end{flushleft}


\end{document}